\documentclass[aps,prd,preprint,groupedaddress, showpacs, amsfonts,amssymb,amsmath, floatfix]{revtex4}

\usepackage{bm}
\usepackage{graphicx}

\begin{document}

\title{A Crystal-based Matter-wave Interferometric Gravitational-wave
Observatory}

\author{Raymond Y. Chiao}

\email{chiao@physics.berkeley.edu}

\homepage{physics.berkeley.edu/research/chiao/}
{}

\affiliation{Department of Physics, University of California,
Berkeley, CA 94720-7300}

\author{A.~D.~Speliotopoulos}

\email{adspelio@uclink.berkeley.edu}

\affiliation{Department of Physics, University of California,
Berkeley, CA 94720-7300}

\date{December 26, 2003}

\begin{abstract}
It is shown that atom interferometry allows
for the construction of MIGO, the \textit{M}atter-wave
\textit{I}nterferometric \textit{G}ravitational-wave
\textit{O}bservatory. MIGOs of the same sensitivity as LIGO or
LISA are expected to be orders of magnitude smaller than either
one. A design for MIGO using crystalline diffraction gratings is
introduced, and its sensitivity is calculated.
\end{abstract}


\maketitle

\section{Introduction}

Roughly speaking, since an atom with mass $m$ ``weighs'' much more
than a photon with frequency $\omega$, all other things being
equal, matter-wave interferometers are expected to be more
sensitive to inertial effects than light-wave interferometers by a
factor of $mc^2/\hbar \omega\sim 10^{10}$ if atoms are
used \cite{Scully}. Because of this factor, it is expected that
observatories constructed from matter-wave interferometers will be
much more sensitive to gravitational waves (GWs) than those
constructed with light-based interferometers. This expected
increase in sensitivity makes use of the quantum phase of an
atomic deBroglie matter wave $\lambda_{dB} = 2\pi\hbar/mv$ where
$v<< c$ is the speed of an nonrelativistic object with mass $m$.
When a GW passes through an atom interferometer, the speeds of the
atoms in the interferometer\textemdash and hence their deBroglie
wavelengths\textemdash will be slightly perturbed. By suitably
designing the interferometer, the shifts in the phase of the atoms
caused by this perturbation in their deBroglie wavelength will
result in a shift in the interference pattern, which can then be
measured interferometrically. Like LIGO, this shift can be used
to determine the properties of the GW. We shall argue in this
paper that using matter-wave interferometry, a
\textit{M}atter-wave \textit{I}nterferometric
\textit{G}ravitational-wave \textit{O}bservatory (MIGO) can be
constructed. For the same sensitivity, MIGO is expected to not
only be orders of magnitude smaller than laser-based observatories
such as LIGO (\textit{L}aser \textit{I}nterferometer
\textit{G}ravitational-wave \textit{O}bservatory) and LISA
(\textit{L}aser \textit{I}nterferometer \textit{S}pace
\textit{A}ntenna), but is also expected to have a frequency
response range much broader than current GW observatories.

While classical-based systems such as LIGO and LISA place \textit{classical}
test masses (mirrors) a certain distance way from the central beam
splitter of the interferometer, and measures slight shifts in their
\textit{positions} through light-based interferometry, quantum-based
systems such as MIGO throw out \textit{quantum} test masses (atoms) with
certain velocities, and measures the quantum phase shift due to slight
shifts in their \textit{speeds} through atom-based interferometry. As
we will show below, given a MIGO with a characteristic length $L_{MIGO}$,
for the same shot-noise-limited phase shifts as LIGO at 125 Hz
(where LIGO is most sensitive)
\begin{equation}
L_{MIGO}\approx\left\{B\left(  \frac{\hbar\omega_{\gamma}}{mc^{2}}\right)
\frac{c}{2\pi fL_{LIGO}}\right\}  ^{1/2}\>L_{LIGO},
\label{MIGO-LIGO-Comp}
\end{equation}
where $2B=150$ is the number of times light bounces back and forth in
each of LIGO's arms, $\omega_\gamma$ is the frequency of the laser used
in LIGO, $m$ is the mass of the atom used in MIGO, $f$ is the
frequency of the GW, and $L_{MIGO}$ is the length of one of LIGO's
arms. Thus, $L_{MIGO}\approx1$ m as compared to $L_{LIGO}=4$ km if a
cesium atom is used. Similarly, a comparison, for equal shot-noise-limited
phase shift, of a space-based MIGO with LISA leads to a MIGO
configuration that at least ten thousand times smaller than LISA.

In this paper, we present one possible configuration for MIGO,
and calculate the sensitivity of this configuration to GWs. We also
outline the technologies that could potentially be used to construct
it. Even though we restrict ourselves to those technologies that have
already been demonstrated in various atom diffraction and
interferometry experiments, our purpose in this paper is to explore
the extreme limits of these technologies, and through these limits,
establish the fundamental limitations to MIGO's capabilities. In this
vein, we are particularly interested in crystalline diffraction
gratings and mirrors.

While the diffraction of helium atoms off of cleaved lithium
fluoride crystals was demonstrated by Otto Stern over seven
decades ago \cite{Stern}, and single-crystal Bonse-Hart neutron
interferometers was used to measure gravitational
effects \cite{COW} three decades ago, atom interferometers using
the natural periodic structure of crystals as reflective
diffraction gratings, i.~e., crystalline gratings, to make beam
splitters, have yet to be demonstrated. We will nevertheless
consider only crystalline gratings in this paper. We shall show
that MIGO's overall length depends crucially on the periodicity of
the diffraction grating used in the beam splitter; the smaller the
periodicity, the shorter the interferometer. While in principle
any periodic structure can be used to construct diffraction
gratings, the smallest of such possible structures is limited by
the natural periodicity of crystals. With a periodicity on the
order of \AA ngstr\"oms, diffraction gratings made from single
crystals will have a periodicity orders of magnitude smaller than
current nanofabricated gratings, whereas diffraction gratings
based on light have a periodicity in the hundreds of nanometer
range. MIGOs constructed with crystalline gratings will thus be
the smallest matter-wave-based GW observatories that can
conceivably be made.

\section{Brief Review of Atom Interferometry}

Matter-wave interferometry is based on the particle-wave duality
principle of quantum mechanics. This principle states that every
massive quantum object can have characteristics of both a particle
and a wave. Which of these two characteristics it exhibits depends
on the properties the experimentalist wishes to measure or
exploit. By exploiting the wave nature of particles, matter-wave
interferometers have been constructed using ``elementary''
particles such as neutrons \cite{COW}, as well as ``composite''
particles such as atoms \cite{Pritchard, Kas1}, and
C$_{60}$ fullerene molecules \cite{Zeilinger}.

Like light-wave interferometry, matter-wave interferometers can be
divided into three distinctive parts: the source emitting the
interfering particle, the ``atom optics'' consisting of
beam-splitters and mirrors, and the detector. Similar to the beams
of light in light-wave interferometers, the beams of particles passing
through the interferometer picks up a phase shift as well, but now
$\Delta\phi=\left(S_{cl}^{\gamma_{1}}-S_{cl}^{\gamma_2}\right)/\hbar$,
and $S_{cl}^{\gamma_{1}}$ and $S_{cl}^{\gamma_2}$ are the action
for a particle travelling along the two \textit{quasi-classical}
paths $\gamma_{1}$ and $\gamma_{2}$ along the two arms of the
interferometer. Large throughputs of atoms will be needed for MIGO,
as well as advanced ``atom optics''; we focus therefore here on
the atomic beam source, and the beam splitters and mirrors, and
refer the reader to Scoles \cite{Scoles} for an review of
atomic-beam detectors.

\subsection{Supersonic Sources and 2D Optical Molasses Collimation}

Supersonic sources were first developed in the 1960s by chemists to
study of chemical reactions \cite{Scoles}, and they have a number of
properties superior to the effusive sources (ovens) which are more
often used by physicists. While the velocity distribution of atom
beams from effusive sources is essentially a Maxwellian one that is
fixed by the temperature of the oven, the velocity spreads of
supersonic sources are much narrower. Fractional velocity spreads of
1\%\textemdash0.3\% for helium\textemdash have been achieved from
continuous supersonic sources, and even smaller fractional velocity
spreads can be achieved using pulsed sources. Nearly monoenergetic
beams with very high intensities can thus be produced. Pritchard
\cite{Pritchard}, for example, has produced $10^{21}$
atoms/cm$^{2}$/s/sr sodium beams, and Toennies \cite{PrivateToennies}
has produced very cold helium beams with $1.5\times10^{19}$ to
$1.5\times10^{20}$ atom/sr/s. {

\bigskip
\noindent\textbf{Continuous Supersonic Sources:}
A typical continuous supersonic source functions as follows: A jet
of gas from a high-pressure reservoir escapes supersonically in
free expansion through a nozzle, consisting of a small orifice
typically 10 to 100 microns in diameter, into a differentially
pumped low-pressure chamber that has a larger orifice at its
output end called the ``skimmer''. This skimmer has the
appropriate geometry to skim away the hotter components of the
outer perimeter of the rapidly expanding gas jet, thus leaving
only the intense, low-temperature central component of the atomic
beam to enter into another differentially pumped chamber.
Importantly, after the expansion the atoms in the beam no longer
collide with one another. The beam is often collimated yet again
using a second slit at the output end of yet another
differentially pumped chamber before it enters the main vacuum
chamber containing the atom-based device, such as an
interferometer \cite{Pritchard}. With successive stages of
differential pumping by means of either diffusion or
turbomolecular pumps, one can maintain an ultra-high vacuum in the
main chamber that is often needed. In addition, using optical
molasses techniques \cite{Zeeman}, beam collimation can be achieved
without throwing away most of the atomic beam, as is currently
done when collimating slits are used. The combination of 2D
optical molasses collimation with effusive sources has been
achieved by Kasevich \cite{Kas1}. However, its combination with
supersonic sources, as we propose here, can yield the
high-brightness atomic beams necessary for good signal-to-noise
ratios in MIGO.

\bigskip
\noindent\textbf{Pulsed Supersonic Sources:}
Most pulsed supersonic sources function in much of the same way as
continuous supersonic sources, but with the addition of a fast-acting
valve to pulse the beam. Duty cycles of roughly 10$^{-4}$ can be
achieved resulting in very high fluences. Because
of the short duty cycles, smaller pumps than those used in continuous
supersonic sources can be used. Of particular interest is the pulse
source developed by Powers et.~al.~\cite{Powers}, which uses laser
ablation of metallic sources such as lithium \cite{LAblate} to generate
the atomic beam. The high-pressure gas chamber is replaced by the
laser-plus-metal assembly, and optical molasses can then be added to
reduce the velocity spread of the pulsed supersonic beam as well.

\subsection{Atomic Beam Optics}

There are a number of approaches that can be taken to fabricate the
``atom beam optics''\textemdash the beam splitters and
mirrors\textemdash needed for the construction of MIGO. These optics
have been constructed by using nanofabricated physical
gratings \cite{Pritchard,Zeilinger}, and by also using
light through either standing waves or stimulated raman
pulses \cite{PCC}. We shall focus on the possibility of using cleaved
surfaces of single crystals, which yield near-perfect crystalline
surfaces, in the construction of optical elements for atomic beams.

Atomic beam mirrors can be constructed either out of diffraction
gratings, or from specular reflection off of smooth
surfaces. Diffraction orders of atomic beams from crystalline
surfaces have a history stretching back to Otto Stern's
experiments \cite{Stern}, and research in this area continues
today \cite{Helium-Nickel}. Consequently, construction of zero-diffraction
order, crystal-based mirrors is certainly possible. Of
more interest for MIGO is the work of Holst and Allison, however. They
have demonstrated that the flexible, thin, hydrogen-passivated
single-crystal silicon wafer surface Si(111)-($1\times1$)H can be used
as a specular mirror for focusing helium atomic beams \cite{Holst2000}. The
development of these mirrors can not only be used to construct
\textit{collimating optics} for the atomic beam, but they also demonstrated
that planar, specular-reflection \textit{atomic mirrors} constructed from
single crystals are possible as well. We shall see that the
sensitivity of MIGO is greater if planar, specular-reflection mirrors
are used instead of diffraction gratings as reflection optics, and we
will focus primarily on these types of mirrors here.

Since the diffraction of atomic beams off of single crystals has been
demonstrated, it is possible, in principle, to construct
\textit{coherent} beam splitters using single-crystal diffraction
gratings that work in reflection. Such diffraction gratings have a
periodicity of a few \AA ngstr\"oms, and will be orders of magnitude
smaller than that of current nanofabricated diffraction
gratings. Alignment of interferometers constructed from crystal
diffraction grating is much more challenging, however, and it is still
an open question as to whether considerations arising from the
Debye-Waller factor will allow for the construction of these
interferometers. We address both issues below.

\bigskip
\noindent\textbf{Laser Interferometer Alignment:}
Since crystal-based atom interferometers must be aligned within an
\AA ngstr\"om or so, laser interferometry must be used to position
and align the various optical elements of these interferometers.
Although this is a difficult engineering task, there is no
fundamental obstacle towards this goal. The sensitivity of the
laser interferometry in distance measurements to make such
measurements arises from the photon-shot noise limit
$\delta\varphi_{laser} \simeq 1/\{\dot{N}_{photon}\tau\}^{1/2}
\simeq 1\text{\AA}/1\mu\text{m}\simeq 10^{-4}$, where
$\dot{N}_{photon}\tau$ is the number of photons in a given
measurement time $\tau$ interval. Hence we would need
$\dot{N}_{photon}\tau \simeq10^{8}$ in order to perform these
measurements. This is not difficult achieve with standard lasers.

The argument in support of the feasibility of aligning
crystal-based atom interferometers is further strengthened when one
realizes what has already been achieved by LIGO. Unlike LIGO,
where the mirrors of the interferometer are in free fall, i.e.,
placed as pendula on the end of kilometer-long interferometer
arms, as we shall see below, the mirrors and beam splitters in
MIGO can be fixed to an underlying, mechanically rigid frame.
Furthermore, the size of MIGO is on the order of meters, and not
of kilometers as in the case of LIGO, or of millions of kilometers
as in the case of LISA. This should make the task of aligning MIGO
orders of magnitude easier than that for LIGO. Indeed, the
active-feedback-mechanism technology developed for LIGO to
stabilize the mirror-assembly\textendash central-beam-splitter
distance to within $10^{-13}$ m could be transferred to MIGO as
well \cite{LIGO-Report}.

\bigskip
\noindent\textbf{Decoherence and the Debye-Waller factor:}
Quantum interference requires that there exists no possibility of
``which-path'' information for the helium atom inside the
interferometer. This requires that diffraction of atomic beams off
of beam splitters, and their reflection off of mirrors be
elastic and coherent. (Scattering of atoms in the beam off of
background gas in the ultra-high vacuum system is assumed to be
negligible). A measure of the inelastic versus elastic components of
these processes is based on the Debye-Waller factor $W$ in the
intensity ratio $I/I_{0}=\exp(-2W)$, where $I$ is the diffracted
(or reflected) intensity and $I_{0}$ is the incident intensity. The
factor $W$ is a measure of the fluctuations of the atoms in the
crystal that diffract the incident helium beam.
For diffracted helium beams, $W=\mathcal{B}/a^{2}$, where $a$ is the lattice
constant of the crystal, and $\mathcal{B} \sim 0.5$
\AA$^{2}$ at room temperature \cite{Debye-Waller}. The rule of
thumb \cite{Scoles} is that $W/12<0.1$, for sharp, elastic diffraction
patterns to be seen. The requirement for observing interference is
more stringent, however; the probability of emitting even a single
phonon during the diffraction process must be negligible,
and $W/12<0.01$ is required \cite{Weare}. For crystal diffraction
grating and mirrors, $W/12\sim 10^{-3}$ at room temperature and
decreases at lower temperatures. Thus it is highly probable that the
\textit{zero-phonon} process is the dominant one, and therefore
quantum phase coherence is expected in the interferometer.
The observation of high-visibility interference fringes for neutrons
in the Bonse-Hart interferometer \cite{COW} is evidence that this is
indeed the case.

\section{LIGO, LISA, and the Detection of GWs}

The great majority of the current experimental searches for GWs are based on
laser interferometry. These detectors are scalable by design to such
sizes that the detection of GWs becomes feasible. A number of research
groups located throughout the world \cite{Robertson} are expecting to
begin to collect data soon: GEO600, German-British collaboration;
VIRGO, a French-Italian collaboration; TAMA300, a Japanese effort; and
ACIGA, an Australian effort. The current US-based, international
collaboration, is LIGO. In addition, a space-based laser
interferometer system LISA is currently in the initial planning
stage. We will focus specifically on LIGO.

LIGO is a set of three interferometers based at two locations separated by
3020 km: Hanford, Washington and Livingston,
Louisiana \cite{LIGO-Report}. All three instruments
are based on Michelson interferometers with Fabry-Perot arms. The physical
length of the each arm of the main LIGO interferometer is 4 km, and with the
Fabry-Perot interferometers in each arm, the optical path of the arm is
increased by the bounce factor, $2B= 150$. At the end of each arm is a massive
mirror suspended vertically as pendula within a vacuum chamber, and
the location of the mirror assembly must be held in position within
$10^{-10}$ to $10^{-13}$ m with respect to the center of the
interferometer.

In its current LIGO I configuration, LIGO is designed to detect GWs in
the frequency range of 40 to 10$^{4}$ Hz. At frequencies below 40 Hz
seismic noise causes a rapid decrease in LIGO's sensitivity, and at
frequencies above 125 Hz the shot-noise limit$^{14}$
begins to gradually limit LIGO's sensitivity as well. Between 40 and
125 Hz LIGO's sensitivity is limited by thermal noise in the
suspension system of the end mirrors. Construction of LIGO I began in
1996, and the main interferometers were commissioned in 2001. The
first science runs were started in June of 2002; data from these runs
are currently being analyzed. Upgrading LIGO I to LIGO II, which is
designed for GW astronomy, is expected to begin in 2006 and to be
completed by 2007.

\section{The \textit{M}atter-wave \textit{I}nterferometric \textit
{G}ravitational-wave \\\textit{O}bservatory (MIGO)}

\subsection{The Vertical MIGO Configuration}

Fig.~$\ref{g-Dominated}$ is a diagram of the vertical MIGO
configuration, one of the possible configurations for MIGO.
We start here with a simplified description of the interferometer. Detailed
analysis of the response of this configuration to the passage of GWs
will be presented below.

In the vertical configuration for MIGO shown in
Fig.~$\ref{g-Dominated}$, alkali atoms are emitted from a pulsed
supersonic source, and are then \textit{slowed down} using Zeeman
slower \cite{Zeeman} to velocities so slow that the acceleration
due to gravity $g$ dominates the trajectories of the atoms through
the interferometer. Immediately after the supersonic source, the
beam is then collimated using 2D optical molasses to further
narrow the transverse velocity spread of the atomic beam. After
the beam is diffracted off the initial beam splitter atoms in the
$n=\pm1$ orders, mirrors, which retroreflect the almost parabolic
upwards trajectories, are placed at the top of these trajectories.
(Atoms in the higher diffraction orders are not used.) The atoms
are then reflected off of these mirrors and fall back downwards,
almost retracing their almost-parabolic upwards trajectories. The
left and right paths of the falling atoms are then recombined by
the same beam splitter, and subsequently detected. The
\textit{bilateral} symmetry of this interferometer along the
vertical line ensures that the gross motion of the atoms on the
left side of the interferometer is reflected on the right side.
The parabolic trajectories of the atoms is shifted slightly during
the passage of a GW, and, as can be seen from the force lines in
Fig.~$\ref{g-Dominated}$, will be slightly different for atoms
travelling along the left hand path compared to the right hand
path. It is this \textit{asymmetrical} shift of the atomic
trajectories by a $\times$-polarized GW that leads to the overall
phase shift.

\subsection{MIGO and LIGO: The Underlying Physics}

In this section, we focus on the physics that underlies both MIGO and LIGO, and
compare their relative sensitivities to GWs from astrophysical
sources. We shall emphasize only the physics here, and our arguments
will be more physical than formal; an exact calculation of the MIGO
phase shift can be found in the next section. As such, in this section
we consider only the effects of a GW passing over freely-falling
particles, and neglect all other external forces on the system.

As is well known, when a GW passes through a system, its effects
on \textit{all parts} of the system \cite{MTW,ADS1995}
must be included. It is \textit{not} possible to shield against
gravitational forces as it is for the other forces of nature. Only
\textit{differences} in the motion between objects\textemdash such
as the mirrors and the beam splitters in an
interferometer\textemdash can  be measured; their absolute motion
cannot. As a consequence, the motion of an object \textit{relative
to an observer} is governed by the geodesic \textit{deviation}
equation \cite{MTW} containing an effective tidal acceleration
acting on the object arising from the local curvature of the
spacetime. In the long wavelength limit, this equation reduces to
\begin{equation}
\frac{d^2x^i}{dt^2}= \frac{1}{2} \frac{d^2h_{ij}}{dt^2}x^j,
\label{geodesicdeviation}
\end{equation}
where by construction the origin of coordinate system is placed at the
observer at the beam splitter. The metric deviations from flat
spacetime $h_{ij}(t)$, i.~e., the amplitude of the GW, is a $3\times
3$ tensor, and not a vector as it is for electromagnetic waves,
although GWs have two polarizations\textemdash the $+$
polarization and the $\times$ polarization\textemdash just like
electromagnetic waves.

The geodesic deviation equation lies at the heart of both LIGO and
MIGO: LIGO on the classical level, and MIGO on the quantum level. For
LIGO, one considers a test mass (a mirror) at a distance $L^{i}$ from
the observer (the central beam splitter), which is taken to be the
origin of the coordinate system. When a GW passes over the
apparatus, it causes small shifts $\xi^{i}$ in this distance so that
since $x^{i}=L^i+\xi^{i}$, where $\vert \xi^{i}\vert  <<\left|
L^{i}\right|$,
\begin{equation}
\frac{d^{2}\xi_{i}}{dt^{2}}\approx \frac{1}{2}
\frac{d^2h_{ij}}{dt^2}L^{j}.
\nonumber
\label{LIGOdeviation}
\end{equation}
To lowest order,
\begin{equation}
\xi_{i}\approx \frac{1}{2}L^{j}
\int_{0}^{t}dt^{\prime}\int_{0}^{t^{\prime}}dt^{\prime\prime}
\frac{d^2h_{ij}}{dt^2}=\frac{1}{2}L^{j}h_{ij},
\nonumber
\label{LIGOsolution}
\end{equation}
where for $t\leq0$, we assume $h_{ij}\equiv0$. Thus, in LIGO one puts a
mass (a mirror) at $t=0$ a certain distance $L^{i}$ away from the
observer, and observes fluctuations in its \textit{position} due to the passage
of a GW. These fluctuations are measured by light-based
interferometry. The freely-falling end mirrors of LIGO are designed to
be these $classical$ test masses, and the phase shift due to changes
in the length of its arms is $\Delta\phi_{LIGO}=2\pi \Delta
L_{LIGO}/\lambda_\gamma$, where $\lambda_\gamma$ is the wavelength of
the laser, and $\Delta L_{LIGO}$ is the optical path length difference
of LIGO induced by the GW. For the two arms of the interferometer this
difference for a $+$ polarization GW with amplitude $h_{+}$ is $\Delta
L_{LIGO} =B L_{LIGO} h_{+}$ where the factor of $B$ accounts for
multiple reflections of the light beam within each arm of LIGO within
the Fabry-Perot interferometers \cite{Thorne}. Thus,
\begin{equation}
\Delta\phi_{LIGO}= \frac{2\pi}{\lambda_\gamma} B L_{LIGO} h_{+},
\label{LIGOPhase}
\end{equation}
and depends only on the amplitude of the GW, and the length
$L_{LIGO}$ of LIGO's arms. Strictly speaking this expression for
$\Delta \phi_{LIGO}$ is only applicable for GWs with frequencies
$\sim 125$ Hz; at 4 km, the arms of LIGO is roughly 1/6 the
wavelength of a GW in the upper end of its frequency response
spectrum, and causality has to be taken into
account \cite{TysonandGiffard}. Although design characteristics of
LIGO such as power recycling and shot-noise modifications by the
Fabry Perot arms will change the sensitivity from this simple
form, it is adequate for our purposes. Note that in LIGO the
photons used in making the measurement are \textit{not} the test
particles being acted on by the GW; only the \textit{end mirrors} of the
interferometer are being acted on.

In contrast with LIGO which places an end mirror at a certain distance
away and measures shifts in its position, in MIGO one throws out a
test mass, i.~e., \textit{an atom}, at $t=0$ with a velocity $V_{i}$
from the origin, 
and observes deviations in its $velocity$, and thus in its
path in spacetime, due to the passage of a GW. To see this, we take
$v_{i}=V_{i}+\beta_{i}$, where $\vert \beta_i\vert  <<\vert
V_{i}\vert$. Then from Eq.~$(\ref{geodesicdeviation})$,
\begin{equation}
\frac{d\beta_{i}}{dt}\approx \frac{1}{2}\frac{d^2h_{ij}}{dt^2} V^{j}t.
\nonumber
\label{MIGOdeviation}
\end{equation}
Once again to lowest order,
\begin{equation}
\beta_{i}\approx \frac{1}{2} V^{j} \int_{0}^{t}t'
dt^{\prime}\frac{d^2h_{ij}}{dt^2}\sim \frac{1}{2}V^{j}
t\dot{h}_{ij}
\nonumber
\label{MIGOsolution}
\end{equation}
where $\dot{h}_{ij}$ is the rate of change of the strain field
$h_{ij}^{GW}$ of the GW. The parallelism between the displacement
$\xi^{i}$ in LIGO, and the velocity shift $\beta_{i}$ in MIGO, is
readily apparent. Classically, the velocity shift $\beta_{i}$ is
extremely small, and virtually impossible to measure. Quantum
mechanically, however, this velocity shift can be measured by means of
quantum interference.

\begin{figure}[ptb]
\begin{center}
\includegraphics[width=0.8\textwidth]{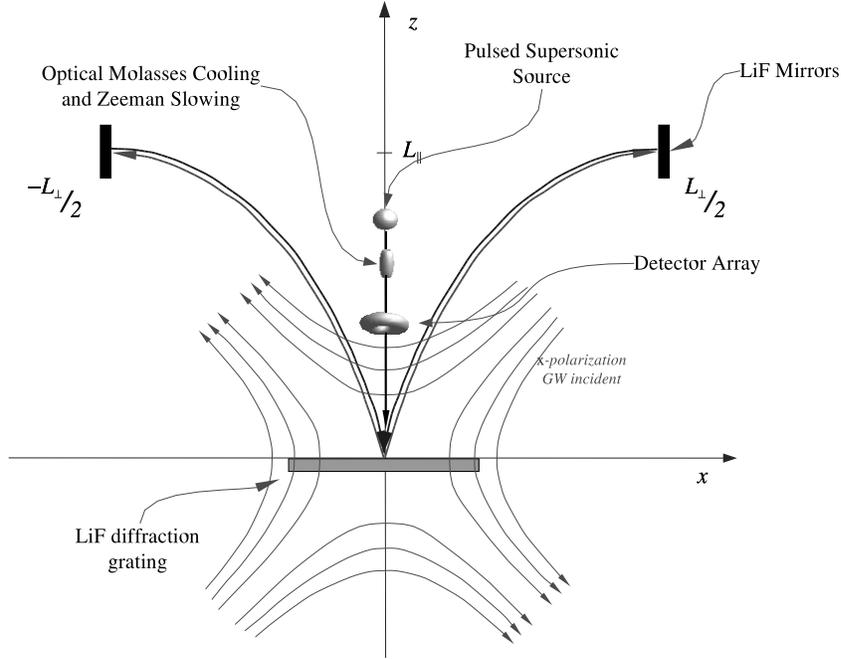}
\end{center}
\caption{Schematic of vertical MIGO configuration, with a $\times$
polarization GW
  incident normal to the plane of the interferometer. }
\label{g-Dominated}
\end{figure}

A quantum particle such as an atom in a matter-wave interferometer
has a deBroglie wavelength $\lambda_{dB}=2\pi\hbar/mv$, where $m$ is
the mass of the atom and $v$ is its speed. Changes in the atom's speed
$\Delta v$ due to GWs result in changes to its local deBroglie
wavelength, and
hence its quantum phase. If $T$ is the time it takes the atom to
traverse MIGO, and if $L_{MIGO}$ is the characteristic size of MIGO,
then $\vert V_i\vert \sim L_{MIGO}/T$, and the shift in the atoms
position $\Delta L_{MIGO}$ after it traverses the interferometer is
$\Delta L_{MIGO}\sim L_{MIGO}T\dot{h}_\times/2$ for a $\times$
polarization GW with amplitude $h_\times$. Like LIGO, this results in
a phase shift, but now of the form
\begin{equation}
\Delta\phi_{MIGO}\sim\frac{4\pi}{\lambda_{\bot dB}}\Delta L_{MIGO}
=2\pi\frac{m}{\hbar} L_{MIGO}^2 f h_\times
\label{MIGOphase}
\end{equation}
where $f$ is the frequency of the GW, and the \textit{deBroglie}
wavelength of the atom is now used instead of the wavelength of
light. Thus, in contrast to LIGO, MIGO's phase shift
$\Delta\phi_{MIGO}$ is proportional to the
\textit{rate of change} of the GW amplitude. Although simplistic, this
rough derivation of the MIGO phase
shift nevertheless elucidates the underlying physics of MIGO. The
above expression for the MIGO phase shift is close to the exact
equation we shall derive in the next
section. Equation $(\ref{MIGO-LIGO-Comp})$ follows directly from
Eqs.~$(\ref{LIGOPhase})$ and $(\ref{MIGOphase})$ by setting
$\Delta\phi_{MIGO} =\Delta\phi_{LIGO}$, and the amplitudes of the two
polarizations of the GW equal.

The relative sizes of MIGO compared with LIGO given in
Eq.~$(\ref{MIGO-LIGO-Comp})$ would seem to be counterintuitive. MIGO
makes use of slowly-moving, nonrelativistic atoms to make its
measurements, while LIGO would seem to make use of fast-moving
photons. At first glance it would seem that MIGO should be
\textit{less} sensitive to GWs than LIGO by some power of
$v_{atom}/c$. This, however, would be an erroneous argument. As
outlined in the above, it is \textit{not} the effect of the GW on the
\textit{photons} used in the laser interferometer that is being
measured in LIGO; it is the effect of the GW on the test masses
(the~\textit{mirrors}) \textit{which are at rest} that is being
measured.

\subsection{Phase Shift Calculation}
In this section we present an explicit calculation of the vertical
MIGO phase shift. As shown in Fig.~$\ref{g-Dominated}$, the origin
of our coordinate system is place at the base of the beam
splitter. In the absence of a GW the atoms travel along parabolic
trajectories. If a GW with $h_{ij}(t)$ is incident along the
$y$-axis, these parabolic paths will be slightly perturbed, and we
can calculate the resulting phase shifts for the atoms by taking
$\vec{x}(t) = \vec{x}_0(t) + \vec{x}_1(t)$ where $\vec{x}_0(t)$ is
the path of the atoms in the absence of a GW, and $\vec{x}_1(t)$
is the first-order perturbation to this path.

If the $r$th atom is sent through the beam splitter at $t=t_r$, the
path of the atom in the $n=+1$ diffraction order in the absence of a
GW is: $v_{0x} = v_\bot$, $v_{0z} = v_\| - g (t-t_r)$, $x_0 = v_\bot
(t-t_r)$, and $z_0 = v_\|(t-t_r) -  g (t-t_r)^2/2$ for
$t_r<t<t_r+ T/2$, where $T/2$ is the time for the atom to travel from
the beam splitter to the mirror, and $g$ is the local acceleration due
to Earth's gravity. Since our focus is on GWs, we approximate $g$ as a
constant, and neglect local curvature effects from stationary sources
such as the Earth. For $t_r+T/2<t<t_r+T$. during which the atom
travels from the mirror back to the beam splitter, its path is $v_{0x}
= -v_\bot$, $v_{0z} = v_\| - g (t-t_r)$, $x_0 = L_\bot - v_\bot
(t-t_r)$, and $z_0 = v_\|(t-t_r) - g (t-t_r)^2/2$, where $L_\bot =
v_\bot T$ is the width of the interferometer and $L_\| = gL^2_\bot /
8v^2_\bot$ is its height. The velocities $v_\bot$ and $v_\|$ are the
initial velocities of the atom as it leaves the diffraction-grating
beam splitter, and $v_\bot = 2\pi\hbar/ma$ where $a$ is the
periodicity of the grating; $v_\| = (v_s^2-v_\bot^2)^{1/2}$ where
$v_s$ is the velocity of the beam incident the beam splitter. A
similar set of equations hold for the $n=-1$ order with
$v_\bot\to-v_\bot$.

The perturbed path $\vec{x}_1$ satisfies the geodesic deviation
equations of motion in Eq.~(\ref{geodesicdeviation}), but with
 $\vec{x}$ approximated as $\vec{x}_0$ on the right hand side. Like
 the unperturbed system, we solve the equations of motion for
 $\vec{x}_1$ by dividing the trajectory of the atom into two parts:
 first from the beam splitter to then mirrors, and then from the mirrors
 back to the beam splitter. The effect of the mirrors on the paths
 of the atoms are taken into account through the mirror boundary conditions
 $v_{1z}(t_r+ T^{-}/2) = v_{1z}(t_r+ T^{+}/2)$ and $v_{1x}(t_r+
 T^{-}/2) = -v_{1x}(t_r+ T^{+}/2)$; the mirrors are taken to
 be rigidly mounted on the frame of the interferometer. The $-$ and
 $+$ superscripts denote the atom approaching and leaving the mirror,
 respectively. This $jump$ boundary condition on the velocity of the
 atoms along  the $x$-axis is due to the instantaneous, impulsive
 force the mirror exerts on the atom when it hits the mirror.

The action for the geodesic deviation equation can be written as
\begin{equation}
S = m\int dt \left( \frac{1}{2} v_i v^i - \frac{1}{2}
\dot{h}_{ij}x^iv^j- gz\right)
\nonumber
\label{action}
\end{equation}
Writing $v^i =  v_0^i+v_1^i$, we find that for the
$n=+1$ path the change in the atom's action $S_{+1}'$ due to the GW
relative to the unperturbed action is \cite{ADS1995}
\begin{eqnarray}
\frac{S_{+1}'(t_r)}{m} &\approx& \frac{1}{2} L_\bot\left[v_{1x}(t_r+T^{-}) -
v_{1x}(t_r+T^{+})\right] + \frac{1}{2}L_\bot v_\bot h_{xx}(t_r+T/2)
\nonumber
\\
&&+ L_\| v_\bot h_{xz}(t_r+T/2)  -
\frac{1}{2}\int_{t_r}^{t_r+T}dt\left\{
h_{xx}(t)v_\bot^2+h_{zz}(t)v_{0z}^2(t)\right\}
\nonumber
\\
&{}&
- v_\bot\left\{\int_{t_r}^{t_r+T/2}dt\> h_{xz}(t)
v_{0z}(t)-\int_{t_r+T/2}^{t_r+T}dt\> h_{xz}(t) v_{0z}(t) \right\}
\nonumber
\\
&{}& +\frac{1}{2} g \int_{t_r}^{t_r+T}dt\> \left[h_{xz}(t) x_0(t) +
h_{zz}(t) z_0(t)- 2z_1(t)\right],
\nonumber
\label{S-n+1}
\end{eqnarray}
after successive integration by parts, and using the geodesic
deviation equation of motion. For the $n=-1$ path, we get
$S'_{-1}(t_r)$ from $S'_{+1}(t_r)$ by taking $v_\bot\to-
v_\bot$ in the above. Then,
\begin{eqnarray}
\frac{\hbar}{m} \Delta\phi_{MIGO} &\approx&
L_\bot\left[v_{1x}^{+1}(t_r+T^{-}/2) + v_{1x}^{-1}(t_r+T^{-}/2)\right]
+
\nonumber
\\
&{}&
2 v_\bot L_\| h_{xz}(t_r+T/2)
-
2v_\bot\Bigg\{\int_{t_r}^{t_r+T/2}dt\> h_{xz}(t) v_{0z}(t)
\nonumber
\\
&{}&
-\int_{t_r+T/2}^{t_r+T}dt\> h_{xz}(t) v_{0z}(t) \Bigg\}
+ g \int_{t_r}^{t_r+T} dt\> h_{xz} x_0(t)
\nonumber
\\
&{}&
- g\int_{t_r}^{t_r+T} dt\> \left[z_1^{+1}(t) -z_1^{-1}(t)\right],
\nonumber
\label{PhiMIGO}
\end{eqnarray}
where the superscripts $\pm1$ denote the trajectories for the $\pm1$
orders, and we have used the fact that $x^{+1}_0(t) = -x^{-1}_0(t)$,
while $z_0^{+1}(t) =
z_0^{-1}(t)$. Then,
\begin{equation}
(v_{1x}^{+1}+ v_{1x}^{-1})(t_r+T^{-}/2)=L_\|
\dot{h}_{xz}(t_r+T/2)+ v_\| h_{xz}(t_r) - g\int_{t_r}^{t_r+T/2} dt\>
h_{xz}(t),
\nonumber
\label{z-path1}
\end{equation}
from Eq.~$(\ref{geodesicdeviation})$, while for $t_r<t<t_r+T/2$,
\begin{equation}
(z_1^{+1}-z_1^{-1})(t) = x_0(t)\left[h_{xz}(t) + h_{xz}(t_r)\right] -
  2v_\bot \int_{t_r}^{t} dt'\>h_{xz}(t'),
\nonumber
\end{equation}
and for $t_r+T/2<t<t_r+T$,
\nonumber
\begin{eqnarray}
(z_1^{+1}-z_1^{-1})(t) &=& L_\bot h_{xz}(t) -v_\bot [t-t_r]
  \left\{h_{xz}(t) -h_{xz}(t_r)\right\}
\nonumber
\\
&{}&
-2v_\bot
  h_{xz}(t_r+T/2)[t-t_r-T/2]
\nonumber
\\
&{}&
+
2v_\bot \left(\int_{t_r+T/2}^{t}
  dt'\>h_{xz}(t')-\int_{t_r}^{t_r+T/2}
  dt'\>h_{xz}(t')\right).
\nonumber
\label{z-path2}
\end{eqnarray}
Integrating, and then taking the Fourier transform, we get
\begin{equation}
\Delta\phi_{MIGO}(f) =-2\pi\frac{m}{\hbar} L_{\bot}L_\| ifh_{\times}(f)
e^{-i\pi fT} F_v(fT/4),
\nonumber
\label{MIGO-g-Phase}
\end{equation}
for a GW with amplitude $h_{\times}(f)$ and frequency $f$. Only the
$\times$ polarization causes a net phase shift in the interferometer;
the $+$ polarization does not, due to the
bilateral symmetry of the interferometer. The resonance factor is 
\begin{equation}
F_v(fT/4) = 1-\{\hbox{sinc}(\pi fT/2)\}^2+\frac{i}{\pi
 fT/2}\left\{1-\hbox{sinc}(\pi fT)\right\}
\nonumber
\label{resonance}
\end{equation}
where $\hbox{sinc}(x) \equiv \sin{x}/x$. Note that $F_v\to 1$ as $fT\to
\infty$, and thus MIGO is most sensitive to GW in the high frequency
domain. This sensitivity is a direct result of the specular reflection
mirrors, and their effects on the atoms are embodied by the jump
boundary conditions that we used. If diffraction-grating mirrors were
used instead, the high-frequency sensitivity of MIGO would be
drastically reduced.

While $\Delta\phi_{MIGO} \propto L_\bot L_\|$, because the
acceleration due to gravity, $L_\| = gL_\bot^2/8 v_\bot^2$, where
$v_\bot$, the horizontal velocity of atom after the beam splitter, is
inversely proportional to the periodicity of the diffraction grating
used. In particular, we see that $L_\|\propto a^2$; the smaller the
periodicity of the diffraction grating, the shorter the
interferometer; hence the use of crystalline diffraction gratings. We
thus see $\Delta\phi_{MIGO}$ for the vertical MIGO configuration is
proportional to $L_\bot^3$, and not to $L_{\bot}$ as it is for
LIGO. Indeed, taking the absolute value of $\Delta\phi_{MIGO}$, the
shot-noise-limited sensitivity for the vertical MIGO is
\begin{equation}
\tilde{h}(f)_{shot}^{MIGO}=\frac{\hbar}{2\pi
  mL_\bot L_\| f|F_v(fT/4)|\dot{N}^{1/2}}
\nonumber
\label{MIGO-Sen}
\end{equation}
where $\dot{N}$ is the number of atoms passing though the
interferometer per second. In comparison,
$\tilde{h}(f)_{shot}^{LIGO}=\left\{\hbar\omega_{\gamma}/2I_o\eta
c\right\}^{1/2}(2\pi f/\omega_{\gamma})$, where $I_{o}$ is the
power of the laser, and $\eta$ is the photodetector efficiency
(Eq.~123a of Thorne \cite{Thorne}). Thus, while
$\tilde{h}(f)_{shot}^{LIGO}$ \textit{decreases} at higher
frequencies, $\tilde{h}(f)_{shot}^{MIGO}$ \textit{increases}. This
complementarity between the two sensitivities is due to
fundamental differences in the signals measured by LIGO and MIGO:
LIGO measures \textit{position}, while MIGO measures
\textit{velocity}.

\begin{figure}[ptb]
\begin{center}
\includegraphics[angle=270, width=\textwidth]{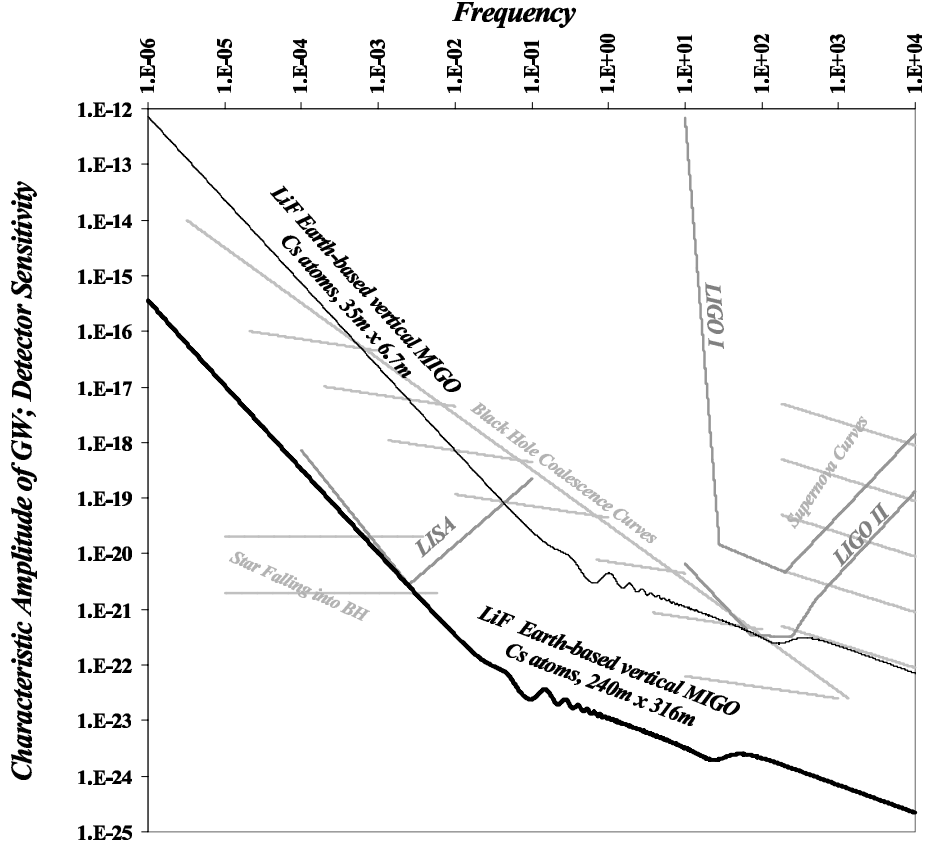}
\end{center}
\caption{The ability of MIGO to detect GWs from various classes of
 burst sources is plotted, and compared with that of LIGO I, LIGO II,
 and LISA.}
\label{h-Burst}
\end{figure}

In the above analysis the mirrors of the interferometer is assumed to
be rigidly attached an infinitely-rigid interferometer that does not
move when the GW passes through. This is an approximation. A more
accurate model is to attach each mirror to a spring with a resonance
frequency $f_0$, which depends on the size of the interferometer as
well as its material properties, and a quality factor $Q$. When the GW
passes through the interferometer it will induce a velocity shift in
the mirror, which will in turn induce an additional shift in
the atom's velocity. The net affect is to subtract the
term $-f^2/(f^2-f_0^2+if_0 f/Q)/2$ from $F_v$. The small dips at the
high-frequency end of the MIGO graphs shown in Fig.~$\ref{h-Burst}$
are due to the shift in the mirror's position by the GW where a speed
of sound of $6000$ m/s was taken for the material used in constructing
MIGO, and the mounting of the mirror to the frame has a $Q=1$.

\subsection{Potential Sensitivity}

The graphs for MIGO are plotted using $\tilde{h}_{MIGO}$, and
$\dot{N}=10^{18}$ atoms per second. An atomic beam of $^{133}$Cs
atoms from a 2000 K pulsed supersonic source is slowed down using
a Zeeman slower, and additional 2D optical cooling is used to
narrow the velocity spread of the atoms perpendicular to the beam.
A LiF crystal, which has a lattice constant of $2$ \AA, is used
reflection diffraction grating for the beam splitter; the mirrors
are constructed out of single LiF crystals as well. Two sizes of
MIGOs are graphed. The first has a width $L_{\bot}=35$ m and a
height $L_{\Vert}=6.7$ m; the atoms in the beam of this MIGO must
be slowed by a factor of 69 by the Zeeman slower. The second has a
width $L_{\bot}=240$ m and a height $L_{\Vert}=316$ m; the atoms
in the beam must be slowed by a factor of 10 by the Zeeman slower
in this case. Graphs of GW signals from burst sources in the
figure were replotted from Thorne's \cite{Thorne} Fig.~9.4. The
plots LIGO I, LIGO II and LISA have been updated from Thorne's
original figure, and were generated from Thorne \cite{Thorne},
Fritschel, and Cutler \cite{Fritschel}, respectively, using Eq.~111
of Thorne \cite{Thorne}.

At $35$ m $\times$ $6.7$ m, the smaller of the two MIGO configurations
graphed has the same sensitivity as LIGO II, in LIGO II's operating
regime, while being a very small fraction of its size. Similarly, at
$250$ m $\times$ $316$ km, the larger of the two MIGO configurations
has the same sensitivity as LISA in LISA's operating regime, even
though its length is only one ten-thousandths of LISA's length of 5
million km. Importantly, in contrast to LISA, this MIGO is earth-based
rather than space-based. Both MIGO configurations are sensitive to a
wider range of GW frequencies than LIGO and LISA, and they would
extend the reach of GW observatories into frequency ranges\textemdash
in particular between $0.1$ and 10 Hz\textemdash not currently
accessible by either LIGO or LISA. Note also the sensitivity
to high-frequency GWs of all three MIGO configurations graphed in
Fig.~3, compared with the gradual \textit{loss} of sensitivity of LIGO
at frequencies above 125 Hz, and of LISA at frequencies above
$2\times10^{-2}$ Hz. This is due to differences in the signal being
measured by matter-wave- and by light-wave-based interferometers: MIGO
measures shifts in the atom's \textit{velocity}, and thus the
\textit{rate of change} of the GW amplitude with time; the larger the
velocity shift, the larger the phase shift. This rate of change
\textit{increases} at higher frequencies, and is the underlying reason
why the smallest-amplitude GW that the three MIGO configurations can
detect in Fig.~3 \textit{decreases} as $1/f^{1/2}$ at high
frequencies. In contrast, LIGO (and LISA) measures the
\textit{position} of the end mirrors, and depends only on the
amplitude of the GW. Indeed, because a GW is time varying, the
smallest-amplitude GW that LIGO can detect must \textit{increase} as
$f^{1/2}$; it actually increases as $f^{3/2}$ because of shot-noise
limits to $\Delta\phi_{LIGO}$. This difference between MIGO and LIGO
can as be explicitly seen by comparing
\begin{equation}
\frac{\tilde{h}(f)_{shot}^{MIGO}}{\tilde{h}(f)_{shot}^{LIGO}}\approx
\frac{\hbar\omega_{\gamma}}{mc^{2}}\left(\frac{2\dot{N}_{\gamma}\eta}{\dot{N}}
\right)^{1/2}\left(\frac{\lambda_{GW}}{2\pi L_{MIGO}}\right)
^{2}|F_v(fT/4)|^{-1}
\nonumber
\end{equation}
in the shot-noise limit where $\dot{N}_{\gamma}$ is the rate photons
enter LIGO.

\section{Conclusion}
We have shown in this paper that in terms of sensitivity, size,
capability, and flexibility, the \textit{quantum} methods embodied in MIGO
have overwhelming advantages over the \textit{classical} methods embodied
in LIGO in detecting GWs. What before took many kilometers with
LIGO could be done in a couple of meters, and what took
millions of kilometers with LISA could be done with only hundreds
of meters. Most importantly, many of the technologies needed in
constructing a MIGO have already been \textit{separately}
demonstrated. Thus, quantum interference can become an important
tool for study of fundamental problems at the intersection of
quantum mechanics and general relativity \cite{Chiao}.

\section{\noindent{\textbf{Acknowledgements:}}}

\noindent ADS and RYC were supported by a grant from the Office of Naval
Research. We thank Chris McKee, John Garrison, Theodore H\"ansch, Jon Magne
Leinaas, Andrew P.~Mackenzie, Richard Marrus, Joseph Orenstein,
William D. Phillips, and Jan Peter Toennies for many clarifying and
insightful discussions.

\end{document}